\documentclass[amssymb,nobibnotes,showpacs, superscriptaddress, aps, prd]{revtex4}
\usepackage{amsmath}
\usepackage{graphicx}
\begin{document}
\title{Avoided level crossing, correlation and entanglement of  two-component Bose-Einstein condensates in a double well}
\author{Weibin Li}
\email{weibinli@wipm.ac.cn} \affiliation{State Key Laboratory of
Magnetic Resonance and Atomic and Molecular Physics, Wuhan Institute
of Physics and Mathematics, Chinese Academy of Sciences, Wuhan
430071, Peoples Republic of China } \affiliation{Graduate school,
Chinese Academy of Sciences, Beijing 10080, Peoples Republic of
China}
\author{Wenxing Yang}
\affiliation{State Key Laboratory of Magnetic Resonance and Atomic
and Molecular Physics, Wuhan Institute of Physics and Mathematics,
Chinese Academy of Sciences, Wuhan 430071, Peoples Republic of China
} \affiliation{Graduate school, Chinese Academy of Sciences, Beijing
10080, Peoples Republic of China}
\author{Xiaotao Xie}
\affiliation{Department of Physics, Huazhong University of Science
and Technology, Wuhan 430074, People's Republic of China}
\author{Jiahua Li}
\affiliation{Department of Physics, Huazhong University of Science
and Technology, Wuhan 430074, People's Republic of China}
\author{Xiaoxue Yang}
\affiliation{Department of Physics, Huazhong University of Science
and Technology, Wuhan 430074, People's Republic of China}
\begin{abstract}
We consider a novel system of two-component atomic Bose-Einstein
condensate in a double-well potential. Based on the well-known
two-mode approximation,  we demonstrate that there are obvious
avoided level-crossings when both interspecies and intraspecies
interactions of two species are increased. The quantum dynamics of
the system exhibits revised oscillating behaviors compared with a
single component condensate. We also examine the entanglement of two
species. Our numerical calculations show the onset of entanglement
can be signed as a violation of Cauchy-Schwarz inequality of
second-order cross correlation function. Consequently, we use Von
Neumann entropy to quantity the degree of entanglement.
\end{abstract} \pacs{03.75.Gg, 05.30.Jp, 03.75.Lm, 03.75.Mn}
\keywords{condensates, level-crossing, entanglement, correlation}
 \maketitle
\section{introduction}
The study of Bose-Einstein condensates (BECs) immersed in artificial
optical lattices has become one of several focuses of current
interest in the ongoing exploration of the  ultralow temperature
physics \cite{bloch}. Recently, experiments of the superfluid to
Mott-insulator transition \cite{greiner} and the number squeezing
related to quantum optics \cite{orzel} have been realized. They
strikingly open the possibility to study fundamental many-body
physics with highly controllable parameters \cite{bloch,jjj}, as
well as find the potential applications in quantum information
processing \cite{bloch1,jaksch1}. In particular, one may prefer to
study the double well system, for which is considered as  the
simplest many lattice model \cite{jjj,fisher,jacksch}. The essential
underlying physics can be understood within the study of a
double-well BECs with a variable barrier height in the two-mode
approximation \cite{milburn,steel,spekkens} and many rich strongly
correlated quantum properties of condensates, such as squeezing
\cite{steel,javanainen}, the self-trapping of Josephson oscillations
\cite{leggett,raghavan,albiez,williams}, the collapse and revivals
dynamics \cite{robinett}, entanglement \cite{vidal}, and spectra
\cite{kalosakas,wu1} have been examined.

On the other hand, experimental realization of multi-component BECs
\cite{stamper, hall,modugno} has stimulated considerable interests
in the study of storing BECs in lattices when the interspecies
interaction of the mixed multi-component BECs affects the quantum
tunnel and quantum coherence crucially
\cite{crutitsky,ostrovskaya,htng}. Examples include both the
experiment demonstrating of  coherent transport of multi-component
BECs in optical lattices \cite{mandel} and theoretical predications
of coexisting superfluid and Mott phases \cite{chen}, fragmented
condensates with topological excitations \cite{demler}, maximally
entangled atomic states \cite{you}, dimer phases \cite{yip}, and
heteronuclear molecular BECs \cite{damski,moore}. However, the
research of multi-component BECs in the double-well has just begin.
Ashhab et.al. \cite{ashhab} studied the Josephson junction between
two spatially separated condensates with a hyperfine degree of
freedom. They find in the semiclassical limitation  the dynamics of
the system can be described by the Bloch equations \cite{ashhab}.
Recently, Ng and co-workers studied the double-well tunnelling of
two-component BECs \cite{ng} and pointed out the generation of
quantum entanglement of different modes.  As a counterpoint, the
entangling property  in the weak-coupling regime with large number
of atoms is also studied
 \cite{ng1}. It reveals that the atoms of two
components initially separately localized in the different potential
wells can tunnel as entangled pairs when the interspecies
interactions are strong enough. These schemes are of particular
interest for engineering many-particle entanglement in BECs, which
may play a prominent role in the study of quantum computing and
quantum information \cite{ng1}.  Up to now we have only a limited
idea about the properties of two-component BECs in a double-well
compared with the single component system.  In addition, other
aspects, such as spectra and quantum dynamics, have not been
specified. Thus a systematic examination is required.

In this paper, we pay much attention to some significant issues of
the two-component BECs in double-well. At first we put forward an
algebra approach to calculate the eigenenergy and eigenstate
explicitly based on the two-mode approximation. It is simple and can
be coded by a several-line mathematica program. By this, it is
surprised to find the onset of obviously avoided level-crossing
\cite{guhr}. This may have profound impacts on the double-well
systems and may relate certain kind of chaotic behaviors \cite
{guhr}. With the eigenspectra and eigenstate, we study the dynamics
in section III. Rabi oscillation and self-trapping effect
\cite{leggett} are demonstrated with different initial conditions.
In section IV, the correlation and entanglement of the system are
examined especially in two limiting cases \cite{ng,ng1}. We show Von
Neumann entropy are always nonzero whether apparent entanglement is
triggered. However the violation of Cauchy-Schwarz inequality of
second-order cross correlation function can  indicates the entrance
of entanglement. We conclude in section V.

\section{quantum two-mode model}
\subsection{Model Hamiltonian}
The many-body Hamiltonian of atomic BECs in a external potential
$V(r)$, in second quantization, is \cite{leggett}
\begin{eqnarray}\label{h1}
\hat{H}=\int
d\mathbf{r}\hat{\Psi}^\dag(\mathbf{r})\left[-\frac{\hbar^2}{2m}\nabla^2+V(\mathbf{r})\right]\hat{\Psi}(\mathbf{r})
\nonumber\\ +\frac{g}{2}\int
d\mathbf{r}\hat{\Psi}^\dag(\mathbf{r})\hat{\Psi}^\dag(\mathbf{r})\hat{\Psi}(\mathbf{r})\hat{\Psi}(\mathbf{r})
\end{eqnarray}
where $\hat{\Psi}(\mathbf{r})$ and $\hat{\Psi}^\dag(\mathbf{r})$ are
the bosonic annihilation and creation field operators, $m$ is the
particle mass, and $g=(4\pi a_s \hbar^2)/m$, where $a_s$ is the
s-wave scattering length. In studies of double-well BECs, one may
prefer to use the well-known two-mode approximation
\cite{milburn,steel,spekkens} to capture the essential physics. We
study a two-component BECs trapped in a symmetric double-well
potential. The total number of atoms in components A and B of the
condensate are $N_a$ and $N_b$, respectively. With this
approximation and the conservation of the particle number of each
component, the system is described  by the Hamiltonian ($\hbar=1$)
\cite{ng,ng1},
\begin{equation}\label{h2}
\begin{split}
\hat{H}&=\frac{\Omega_a}{2}(\hat{a}_L^{\dag}\hat{a}_R+\hat{a}_R^{\dag}\hat{a}_L)+\frac{\Omega_b}{2}(\hat{b}_L^{\dag}\hat{b}_R+\hat{b}_R^{\dag}\hat{b}_L)+\kappa(\hat{a}_L^{\dag}\hat{a}_L\hat{b}_L^{\dag}\hat{b}_L\\
&+\hat{a}_R^{\dag}\hat{a}_R\hat{b}_R^{\dag}\hat{b}_R)+\frac{\kappa_a}{2}[(\hat{a}_L^{\dag}\hat{a}_L)^2+(\hat{a}_R^{\dag}\hat{a}_R)^2]\\
&+\frac{\kappa_b}{2}[(\hat{b}_L^{\dag}\hat{b}_L)^2+(\hat{b}_R^{\dag}\hat{b}_R)^2]
\end{split}
\end{equation}
where the subscipts $L$ and $R$ denote localized modes in the left
and right potential wells. $\hat{a}_j^\dag(\hat{a}_j)$ and
$\hat{b}_j^\dag(\hat{b}_j)$ are, respectively, the creation
(annihilation) operators of components A and B residing in the $j$th
well, $j=L,R$. The parameters $\Omega_a(\Omega_b)$,
$\kappa_a(\kappa_b)$ and $\kappa$ describe the tunnelling rate,
self-interaction strength of component A (B) and the interspecies
interaction strength.
\subsection{Procedure for the Eigenvalue Equation}
Although there are continuous interests in the study of the
Hamiltonian (\ref{h2}) \cite{ashhab,ng,ng1,chen1}, with the result
that such systems may provide a potential entanglement regime of
macroscopic ensembles, no one seems to have been able to obtain the
eigenspectra systemically even for a small number of atoms
\cite{ng}. In handling the system with nonlinear interactions among
several boson modes, one usually needs to use the well-known Bethe
ansatz \cite{andreev} on the energy eigensates with several
parameters determined by highly complicated reduced system of Bethe
equations. In this section, we shall utilize an efficient and simple
method \cite{wu,wu1} to solve Hamiltonian (\ref{h2}). We show that
the corresponding eigenvalue can be reduced into a differential
equation and thus solved by a simple MATHEMATICA code.

The eigenvalue equation for the Hamiltonian (\ref{h2}),
\begin{equation}
\hat{H}|\Psi_{E,N_a,N_b}\rangle=E|\Psi_{E,N_a,N_b}\rangle
\end{equation}
Denote the energy eigenstates as
\begin{equation}
\begin{split}
|\Psi_{E,N_a,N_b}\rangle=\hat{F}(\hat{a}_L^\dag,\hat{a}_R^\dag;\hat{b}_L^\dag,\hat{b}_R^\dag)|\text{vac}\rangle_A|\text{vac}\rangle_B\\
=\hat{F}(\hat{a}_L^\dag,\hat{a}_R^\dag;\hat{b}_L^\dag,\hat{b}_R^\dag)|\text{vac}\rangle_{A,B}
\ \ \ \ \ \
\end{split}
\end{equation}
where $F$ is a polynomial of the creation operators
$\hat{a}_{L,R}^\dag$ and $\hat{b}_{L,R}^\dag$. The vacuum state
$|\text{vac}\rangle_S=|p=0,q=0\rangle_S$ denotes a Fock state of
species $S$ ($S=A,B$) without any bosons in the left well and right
well. Throughout the following paper, states $|p,q\rangle_S$ denote
Fock states with $p$ atoms of species $S (S=A,B)$ in the left well
and $q$ atoms of species $S (S=A,B)$ in the right well. Noting
$\hat{H}\hat{F}|\text{vac}\rangle_{A,B}=([\hat{H},\hat{F}]+\hat{F}\hat{H})|\text{vac}\rangle_{A,B}$
due to the fact $\hat{H}|\text{vac}\rangle_{A,B}=0$, the eigenvalue
equation becomes
$([\hat{H},\hat{F}]-E\hat{F})|\text{vac}\rangle_{A,B}=0$. Finally,
by using the identities
$\alpha_j|\text{vac}\rangle_{A,B}=0,[\hat{\alpha}_j^\dag,\hat{F}]=0,
[\hat{\alpha}_j,\hat{F}]=\partial{\hat{F}}/\partial{\hat{\alpha}_j^\dag}
(\alpha=a,b) $, it is then straightforward to show that the
polynomial $\hat{F}$ of creation operators $\hat{\alpha}^\dag_{L.R},
(\alpha=a,b; j=L,R)$ satisfies the operator-type differential
equation as follows:
\begin{eqnarray}\label{diff}
\bigg[\frac{\Omega_a}{2}\left(x_1\frac{\partial}{\partial
x_2}+x_2\frac{\partial}{\partial
x_1}\right)+\frac{\Omega_b}{2}\left(y_1\frac{\partial}{\partial y_2}
+y_2\frac{\partial}{\partial y_1}\right)\nonumber\\
+\frac{\kappa_a}{2}\left(x_1^2\frac{\partial^2}{\partial
x_1^2}+x_2^2\frac{\partial^2}{\partial x_2^2}\right)
+\frac{\kappa_b}{2}\left(y_1^2\frac{\partial^2}{\partial
y_1^2}+y_2^2\frac{\partial^2}{\partial y_2^2}\right)\\
+\kappa\left(x_1y_1\frac{\partial^2}{\partial x_1\partial
y_1}+x_2y_2\frac{\partial^2}{\partial x_2\partial
y_2}\right)\bigg]F=\lambda F\nonumber
\end{eqnarray}
where we have used the convenient notations
$x_{1,2}\equiv\hat{a}_{L,R}, y_{1,2}\equiv\hat{b}_{L,R}$. And  $F$
is a polynomial of the form
$F(x_1,x_2,y_1,y_2)=\sum_{n=0}^{N_a}\sum_{m=0}^{N_b}c_{n,m}x_1^nx_2^{N_a-n}y_1^my_2^{N_b-m}$.
The parameter $\lambda$ relates to the energy eigenvalues by
\begin{equation}\label{energy}
\lambda=E-\left(\frac{\kappa_a}{2}N_a+\frac{\kappa_b}{2}N_b\right)
\end{equation}

It is important to note that the operator-type differential equation
(\ref{diff}) can formally be regarded as a $c$-number differential
equation owning to the result that all the operators are mutually
commutable with each other and hence can be solved by any ordinary
way \cite{wu,wu1}.

Substituting the $c$-number polynomial $F(x_1,x_2,y_1,y_2)$ into the
differential equation (\ref{diff}), we find the eigenvalues can be
calculated by solving the matrix equation
\begin{equation}\label{matrix}
M \sigma =\lambda \sigma
\end{equation}
where $\sigma$ is a $D$-dimension column vector, $D=(N_a+1)\times
(N_b+1)$ and $M$ is a square matrix of order $D$ with the matrix
elements
$M_{k,l;m,n}=\Omega_a/2[(N_a-k)\delta_{k,l-1}+k\delta_{k,l+1}]\delta_{m,n}+
\Omega_b/2[(N_b-m)\delta_{m,n-1}+m\delta_{m,n+1}]\delta_{k,l}
+\{\kappa_a/2[k(k-1)+(N_a-k)(N_a-k-1)]+\kappa_b/2[m(m-1)+(N_b-m)(N_b-m-1)]
+\kappa[m k+(N_a-k)(N_b-m)]\}\delta_{k,l}\delta_{m,n}$. To obtain
the eigenvalues of vector $\sigma$, we can express the matrix $M$
by a MATHEMATICA code and thus diagonalize it directly
\cite{code}.
\subsection{Avoided Level-crossing}
We have given a systemical procedure to calculate the eigenenergy
and eigenstates in term of parameter $\lambda$ by solving either
analytically or numerically Eq. (\ref{matrix}). For example, one can
obtain explicit results of the simplest case $N_a=N_b=1$, which is
similar to the results in \cite{ng}. Noting here $\lambda$ is
related to eigenenergy by Eq.(\ref{energy}).  We don't plan on
listing these results due to the large number of eigenvalues
($D=(N_a+1)\times (N_b+1)$). One can validate the efficiency of the
method using the code \cite{code}.

\begin{figure}
\includegraphics[width=8.5cm,height=10cm]{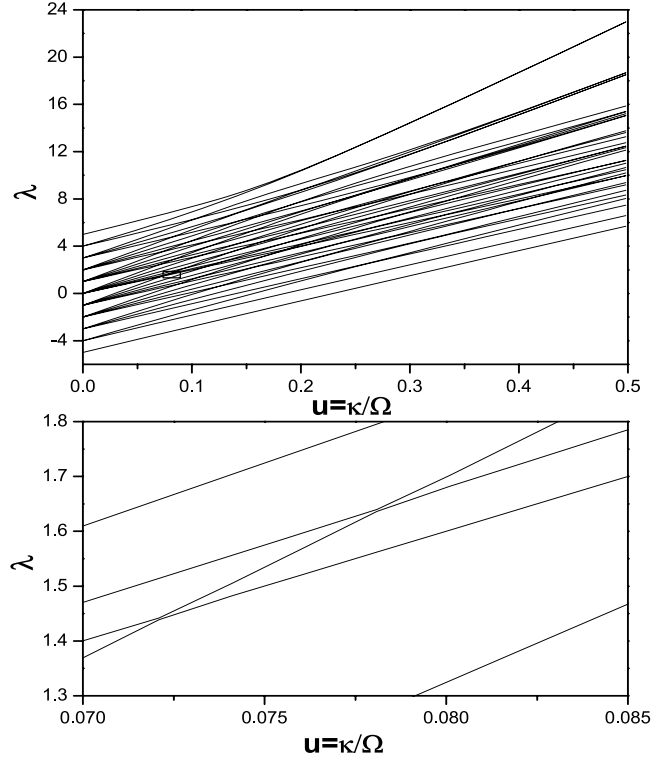}
 \caption{\label{fo}Level for $N_a=N_b=5$. The bottom is the enlargement of rectangle in the top part, which shows avoided level crossing.  }
\end{figure}

The spectra of such systems are usually intricate functions of the
parameters involved in the Hamiltonian (\ref{h2}). To address the
details of spectra for given $N_a$ and $N_b$, we shall plot all of
the spectra. For convenience, we make $\Omega_a=\Omega_b=\Omega$ and
$\kappa_a=\kappa_b=\kappa$. It is a good approximation to $^{87}$Rb
condensate of atoms in hyperfine spin states $|F=2,M_f=1\rangle$ and
$|F=1,m_f=-1\rangle$ with similar scattering lengths \cite{hall}.
Hereafter we  consider only $u\equiv\kappa/\Omega>0$, i.e. the
repulsive situation. All $D=(N_a+1)\times(N_b+1)$ eigenvalues are
depicted in FIG.\ref{fo} (top) for $N_a=N_b=5$. In  cases of
$\Omega\gg\kappa$ and $\Omega\ll\kappa$ most of the eigenstates
remain nearly degenerate. The property of degenerate can be well
understood in the Schwinger representation. Defining
angular-momentum operators of two boson modes of each component as
\begin{eqnarray}\label{sch}
\hat{J}_{\alpha x}=\frac{1}{2}(\hat{\alpha}^\dag_L\hat{\alpha}_L-\hat{\alpha}^\dag_R\hat{\alpha}_R) \nonumber \\
\hat{J}_{\alpha y}=\frac{1}{2i}(\hat{\alpha}^\dag_L\hat{\alpha}_R-\hat{\alpha}^\dag_R\hat{\alpha}_L) \\
\hat{J}_{\alpha
z}=\frac{1}{2}(\hat{\alpha}^\dag_L\hat{\alpha}_R+\hat{\alpha}^\dag_R\hat{\alpha}_L)\nonumber
\end{eqnarray}
where $\alpha=a,b$. Here $\hat{\textbf{J}}_{\alpha}=(\hat{J}_{\alpha
x},\hat{J}_{\alpha y},\hat{J}_{\alpha z})$ are the usual angular
momentum operators. Denote $|j_\alpha,m_\alpha\rangle$  the
eigenstates of $\hat{J}_\alpha^2=\hat{J}_{\alpha
x}^2+\hat{J}_{\alpha y}^2+\hat{J}_{\alpha z}^2$ and $\hat{J}_{\alpha
z}$ with
$\hat{J}_\alpha^2|j_\alpha,m_\alpha\rangle=j_\alpha(j_\alpha+1)|j_\alpha,m_\alpha\rangle$
and $j_{\alpha
z}|j_\alpha,m_\alpha\rangle=m_\alpha|j_\alpha,m_\alpha\rangle$,
where $j_\alpha=N_\alpha/2$. Thus the Hamiltonian (\ref{h2}) may be
written \cite{ng1}
\begin{equation}\label{ha}
\hat{H}=\Omega\hat{J}_z+\kappa\hat{J}^2_x
\end{equation}
where $\hat{\textbf{J}}=\hat{\textbf{J}}_a+\hat{\textbf{J}}_b$ is
the total angular momentum of the system. It is interesting to note
that Hamiltonian (\ref{ha}) looks same as the Hamiltonian of single
component BECs in a double-well \cite{milburn}. The spectra suggests
that for $\Omega$ big the first term in Hamiltonian (\ref{ha})
dominates, in which case the energy eigenstates are close to the
eigenstates of $\hat{J}_z$. Since
$m_\alpha=-j_\alpha,-j_\alpha-1,\cdots j_\alpha-1,j_\alpha$, the
eigenvalues of $j_z$ are $m=-(j_a+j_b),-(j_a+j_b)+1,\cdots
(j_a+j_b)-1,(j_a+j_b)$. Corresponding degeneracies are $1,2,\cdots
(N_a+N_b)/2,(N_a+N_b)/2+1,(N_a+N_b)/2,\cdots 2,1$. This can be seen
obviously in FIG. (\ref{fo}).  While for sufficient big $u$, the
system is dominated by the second term of Hamiltonian (\ref{ha}).
This means that the eigenstates of the Hamiltonian are close to the
eigensates of $\hat{J}_x^2$. From Eq.(\ref{sch}), the operator
$\hat{J}_{\alpha x}$ gives the particle number difference of each
well for one component. So $\hat{J}_x=\hat{J}_{ax}+\hat{J}_{bx}$
will have the eigenvalues $-(j_a+j_b),-(j_a+j_b)+1,\cdots
(j_a+j_b)-1,j_a+j_b$. Hamiltonian (\ref{ha}) is symmetric under
transformation $\hat{J}_x\rightarrow -\hat{J}_x$ which results in
the eigenvalues $0, 1, \cdots (j_a+j_b)^2$ with corresponding
degeneracies $j_a+j_b, (j_a+j_b)-2, \cdots 4, 2$. Although the
complete degenerates are not plotted in FIG. (\ref{fo}), we can note
the trend which demonstrates the constriction of eigenvalues when
$\kappa$ increases. It is important to note the ground state is not
degenerated with respect to $u$ in a large range. This feature means
that the ground state is very stable and hard to be excited to the
excited states.

What appears interesting as the eigenstates undergo a series of
apparent avoided level-crossings in a region of $u$ bounded by the
rectangle. The enlargement plot given in bottom of FIG. (\ref{fo}),
reveals the characteristic double cone structure of the avoided
level crossing. In the intermediate range of $u$, Both the
tunnelling $\Omega$ and the interaction $\kappa$ influence the
Hamiltonian. The competition between them is much complicated. The
authors \cite{wy} have shown the generation of erratic level
crossings, where the system contains only a single component BECs in
a double-well. It is striking that our system shows apparent avoided
level crossings. They are similar to the level repulsion of a
interacting boson model addressed in \cite{arias}, in which the
level repulsion characterizes certain quantum phase transition.
However, it remains to be explored whether the avoided
level-crossings may indicate certain kind of chaotic behaviors or
quantum phase transition. This will be subject of our future
investigations.

\section{quantum dynamics analysis}
With the formal development at hand, we now turn to a discussion of
the quantum dynamics of this system. The prominent properties of
quantum tunnelling dynamics of BECs trapped in a double well depend
on the initial conditions and the relative values of tunnelling rate
and interaction strength \cite{leggett,raghavan,albiez,williams}.
For single-component condensates, many studies
\cite{leggett,raghavan,albiez,williams} have shown that a
self-trapping mechanism can restrict the tunnelling rate
significantly due to the two-body interactions. However, the
dynamics of two-component BECs is effected by two kinds of
interactions, i.e. intraspecies and interspecies interactions.
Although previous study \cite{ng} reveals the paired tunnelling
occurs as interspecies interaction increases, which considers a
situation that the atoms of two components are separately prepared
in different traps initially, a thorough study of quantum dynamics
is not addressed. For example, how the dynamical behaviors of each
component and whole system are influenced by intraspecies and
interspecies interactions is not studied. In the following, by
varying the parameters, we  will examine the quantum dynamics of
such system.

First of all, we determine the quantum dynamics of Hamiltonian
(\ref{ha}), which has the critical relation, $\kappa N=2\Omega$
\cite{milburn} with $N=N_a+N_b$ the total number of particles. It is
very simple and can be constructed by the standard method
\cite{milburn}. Usually, one calculates the Heisenberg equations of
motion
\begin{eqnarray}
\frac{d\hat{J}_x}{dt}&=&-\Omega\hat{J}_y\nonumber \\
\frac{d\hat{J}_y}{dt}&=&\Omega\hat{J}_x-\kappa(\hat{J}_z\hat{J}_x+\hat{J}_x\hat{J}_z)\\
\frac{d\hat{J}_z}{dt}&=&\kappa(\hat{J}_y\hat{J}_x+\hat{J}_x\hat{J}_y)\nonumber
\end{eqnarray}
\begin{figure}
\includegraphics[width=3.15in,height=2.78in]{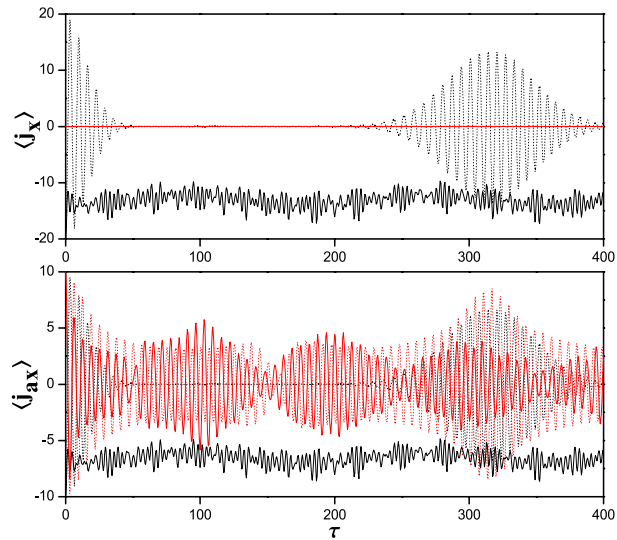}
 \caption{\label{jx20}Dynamics of number differences. The top and the bottom are the mean values of
 $\hat{j}_x$ and $\hat{j}_{ax}$ for two initial conditions $|j,-j\rangle$ (black) and $|j,0\rangle$ (red).
 Here $\Omega=1.0, \tau=\Omega t$ and
$\kappa N=2.5$ for solid curves and $\kappa N=0.8$ for dotted
curves.}
\end{figure}
By representing states of the Hamiltonian (\ref{ha}) in the basis of
$\hat{J}_z$, the time evolution can be obtained by integrating the
Schr\"{o}dinger equation. With the procedure developed in previous
section, this work can also be accomplished immediately. Expressing
the eigenstate of Hamiltonian (\ref{ha}) as
\begin{equation}
|\Psi_{E,N_a,N_b}\rangle=\sum_{n=0}^{N_a}\sum_{m=0}^{N_b}c_{n,m}|n,N_a-n\rangle_A|m,N_b-m\rangle_B
\end{equation}
To study the dynamics of Hamiltonian (\ref{ha}) with a initial
condition $|\Phi(0)\rangle$, we write down the time dependent state
function as
\begin{equation}
\label{time} |\Phi (t) \rangle = e^{ - iHt}|\Phi (0) \rangle
\end{equation}
With the eigenstate $|\Psi_{E,N_a,N_b}\rangle$, we can obtain
\begin{eqnarray}
|\Phi(t)\rangle=\sum_{l=0}^De^{-iE_lt}\langle
\Psi_l|\Phi(0)\rangle|\Psi_l\rangle
\end{eqnarray}
where $|\Psi_l\rangle\equiv|\Psi_{E_l,N_a,N_b}\rangle$, $c_{n,m}$
and  $E_l$ can be obtained by our method numerically.

We take case $N_a=N_b=20$, with $\kappa N/\Omega=2.5$ and $\kappa
N/\Omega=0.8$, corresponding to above and below the threshold. For
simplicity, we make $\Omega=1.0$ and normalize the time in units of
single particle tunneling period, $\tau=\Omega t$ in the numerical
experiments. In FIG. (\ref{jx20}) we plot the mean values of
$\hat{J}_x$ (top), which represents the occupation difference of
 two wells, for two initial states
$|j,-j\rangle\equiv|j_a,-j_a\rangle_A|j_b,-j_b\rangle_B$ and
$|j,0\rangle\equiv|j_a,j_a\rangle_A|j_b,-j_b\rangle_B$ with
$j=j_a+j_b$. The former means the atoms of both components are
prepared in one well initially and the latter corresponds to
separately prepare each component in individual well.  It is not
surprised that the distribution of particles of the latter case is
zero. Thus we focus on examining the evolution with initial state
$|j,-j\rangle$. During long times, the quantum dynamics of
two-component and single component are similar \cite{milburn}. Below
threshold, the dynamics oscillates with revivals at latter times.
While above threshold, the revivals don't appear. It can be regarded
as  the self-trapping \cite{leggett} of two-component BECs. More
interesting, however, is the reversed oscillating properties below
and above threshold compared with the single component ones'
\cite{milburn}. Here the revival and oscillation are quite regular
below threshold while above threshold the mean values are fluctuated
irregularly around constants (depending on the relative values of
$\kappa$, $N$ and $\Omega$).

To obtain more insight of the dynamics, we also plot the number
difference of component A in FIG. (\ref{jx20}) (bottom). It is
obvious the dynamics of each component is similar to the one of
whole system with initial condition $|j,-j\rangle$. The red curves
show that although each component oscillates between two wells
rapidly when the system is separately prepared, the number of atoms
in every well is equal.

On the other hand, when varying $\Omega$ and $\kappa$, the levels
undergo distinguished behaviors, as depicted in FIG (\ref{fo}). One
may prefer to study the dynamics within the regime $\kappa/\Omega\gg
1$ or $\kappa/\Omega\ll 1$. We have done the numerical experiments
in these region and found for the symmetric situations, i.e.
$\Omega_a=\Omega_b=\Omega$ and $\kappa_a=\kappa_b=\kappa$ considered
above, they are trivial. In spite of that, the corresponding
properties of correlation and entanglement are crucial. We will
address these in the following section.

\begin{figure}
\includegraphics[width=3.8in,height=3.0in]{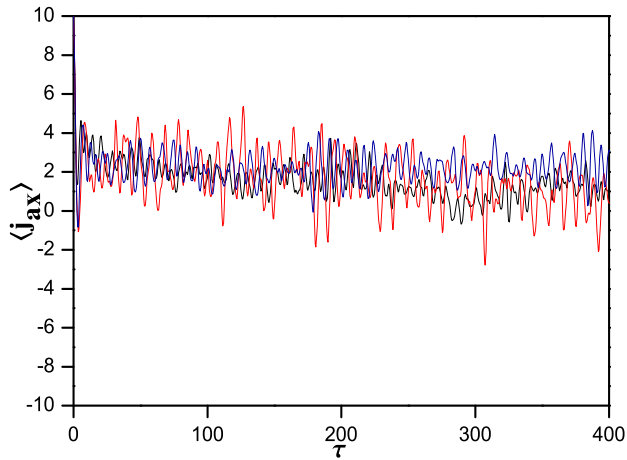}
 \caption{\label{asf}Number differences for component A with initial state $|j,0\rangle=|j_a,j_a\rangle_A|j_b,-j_b\rangle_B$.
 Here black curve represents the case $\Omega=5.3,\kappa=1.2$,
 orange,  $\Omega=1.2,\kappa=5.0$ and blue, $\Omega=0.1,\kappa=4.0$
 .}
\end{figure}

Another interesting phenomenon is self-trapping effect of individual
component. As we have shown, it can  only happen above threshold
when the particles are prepared in a single well initially. Below
threshold or evolving from state $|j,0\rangle$ at $t=0$ it will not
occurs. This is true for the symmetric situation.  However when the
system contains two kinds of atoms (such as $^{41}$K and $^{87}$Rb
\cite{modugno1}) or the interspecies interaction of these
two-component BECs  is varied by the application of magnetic control
\cite{simoni}, this effect will happen for individual components.
Our numerical studies demonstrate  it becomes obvious when the
tunneling $\Omega_a=\Omega_b=\Omega$ are not equal to the
 interactions $\kappa_a=\kappa_b=\kappa$ and at the same
time at least one of them ($\Omega$ and $\kappa$) should be above
threshold. In FIG (\ref{asf}), we have plot some of the results. It
appears that the atoms of each components are favor to stay in the
original wells with time. If we increase the difference between
$\Omega$ and $\kappa$ (both above threshold), the self-trapping of
individual components become more distinct. While one of them below
threshold, it will demand the difference much greater to trigger the
effects.
\section{quantum correlation and entanglement}
A particularly striking aspect of the present study is that it
allows one to obtain quantum correlations between components. The
correlated states of BECs in a double-well model have been a most
active research areas, for which open the opportunity to create
many-particle entanglement in macroscopic systems \cite{you,micheli,
ng,ng1,mahmud}.  You \cite{you}  proposed a scheme to create maximal
entangled cluster states starting from a Mott insulator state.
Micheli et. al. \cite{micheli} studied the possibility to create
dynamically many-particle entanglement of a two-component BECs. They
found within a very short time scale (proportional to 1/N with N the
number of condensate particles), the entangled states are generated.
In particular, Ng et.al.  \cite{ng,ng1} studied the properties of
entanglement with the same model (\ref{h2}). Their studies reveal
that the generation of entanglement can occur under the condition
either (i) the condensates are symmetric with $4\kappa\gg N\Omega$
\cite{ng} or (ii) the condensates are not symmetric with
$\Omega\gg\kappa>\kappa_{a}(\kappa_b)$ \cite{ng1}. While with these
conditions two species of particles are entangled,  what
characterizes the correlations is not clear. On the other hand,
their studies restrict to either strong interaction regime (i) or
weak interaction regime (ii) with a large number of particles.
Studying the system within an expanding parameters range will help
us understand the correlation and entanglement of this system
deeply.
\subsection{Von Neumann Entropy}
At first we introduce Von Neumann entropy \cite{peres} to
characterize the degree of entanglement between these two kinds of
particles. By obtaining the eigenstates and eigenvalues of
Hamiltonian (\ref{h2}), we can calculate the density matrix of the
system at any time $t$
\begin{equation}\label{matrix}
\rho_{AB}(t)=|\Phi(t)\rangle\langle\Phi(t)|
\end{equation}
Then Von Neumann entropy is defined as ($N_a=N_b=N$)
\begin{equation}\label{entropy}
S(t)=-\text{tr}[\rho_A(t)\text{ln}\rho_A(t)]=\text{tr}[\rho_B(t)\text{ln}\rho_B(t)]
\end{equation}
Here $\rho_A(t)$ and $\rho_B(t)$ are reduced density matrices of the
respective subsystems
\begin{equation}\label{submatrix}
\rho_A(t)=\text{tr}_B[\rho_{AB}(t)],
\rho_B(t)=\text{tr}_A[\rho_{AB}(t)]
\end{equation}
Von Neumann entropy describes how much the species are entangled
with time. As a result, we can measure the entanglement of these
states at given parameters and estimate the maximal value of
entanglement of the system. The dimension of the reduced density
matrices $\rho_A$ (or $\rho_B$) is $N_A+1$ (or $N_B+1$). When the
states of the subsystem have equal probability to be occupied, Von
Neumann entropy reaches its maximal value $\text{ln}(N_A+1) \text{or
ln}(N_B+1)$.
\begin{figure}
\includegraphics[width=3.4in,height=3.0in]{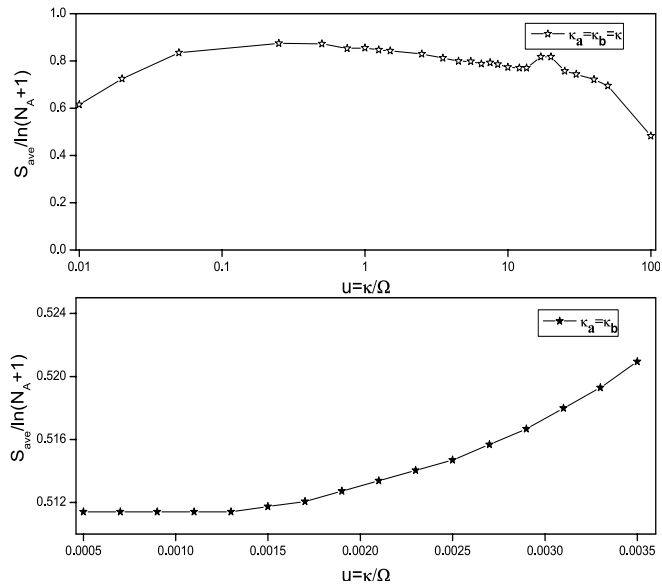}
 \caption{\label{save} Normalized entropy  for different
 $u=\Omega/\kappa$ with $N_a=N_b=20,\Omega_a=\Omega_b=\Omega=1$. Here we plot normalized entropy
 $S_{ave}/\text{ln}(N_a+1)$ of component A. We set  $\kappa_a=\kappa_b=\kappa$ (top) and $\kappa_a=\kappa_b=0.01$ (bottom) accordingly .}
\end{figure}
In order to condense information about the system's entanglement
with various parameters, we characterize Von Neumann entropy by its
time average
\begin{equation}
S_{ave}=\frac{1}{\Delta t}\int_0^{\Delta t} dt S(t)
\end{equation}
employing the averaging interval $\Delta t=30\Omega t$, which
ensures the entropy reaches its saturating values. We start with a
Fock state $|j,0\rangle\equiv|j_a,j_a\rangle_A|j_b,-j_b\rangle_B$,
which denotes component A and component B are prepared in right well
and left well correspondingly. FIG. (\ref{save}) depicts results of
such calculations. The averaged entropy continuously increase with
$u$ until to a maximal value. Then the value of the entropy will not
increase but descend slowly. Note in the parameter range we
considered here the entropy never reaches to $\text{ln}(N_a+1)$. It
is interesting  there is a $jut$ in the descending region. An
approximated calculation \cite{ng} indicates that states of the form
$|n,N_a-n\rangle_A|m,N_b-m\rangle_B$ dominates the dynamics due to
entanglement as $4\kappa\gg N\Omega$.
 However, data in FIG. (\ref{save})
seem to imply particles of two species are entangled adequately even
when $4\kappa\ll N\Omega$. Since one  prefers to paired tunneling as
entanglement, we carefully check the states involved in the
dynamics. When $4\kappa\gg N\Omega$, the system is likely to stay
the states similar to $|n,N_a-n\rangle_A|m,N_b-m\rangle_B$ while
much more other states are excited into the dynamics when
$4\kappa\ll N\Omega$. We also study the situation that
$\kappa_a=\kappa_b\neq \kappa$ while keep the system in the strong
interaction regime. The results indicate that slight difference
between $\kappa_a(=\kappa_b)$ and $\kappa$ will destroy the
entanglement of two species. The system collapses to almost a single
basis state and consequently Von Neumann entropy oscillates within a
small amplitude.

In the so called weak interaction regime \cite{ng1}, we find in a
small range Von Neumann entropy, whether $\kappa>\kappa_a=\kappa_b$
or $\kappa<\kappa_a=\kappa_b$, is larger than the case
$\kappa=\kappa_a=\kappa_b$. However, the numerical study
demonstrates paired tunneling is the major form of  the dynamics
only when $\kappa>\kappa_a=\kappa_b$ within a small range. Exceeding
this range, Von Neumann entropy comes down. See FIG.
\ref{save}(bottom).

\subsection{Quantum Correlation}
Although Von Neumann entropy can be used to quantify the degree of
entanglement of two species, it is not sufficient to estimate how
the system is entangled. We have shown paired tunneling entanglement
and chaotic superposition of states can both contribute to Von
Neumann entropy (FIG. \ref{save}). But what can be used to
distinguish entanglement and chaotic superposition of states has not
been addressed.

Here we calculate the second-order correlation function
\cite{walls94} and demonstrate the onset of paired entanglement
accompanied by the violation of Cauchy-Schwarz inequality. The
single-time two-mode second-order cross-correlation function is
bound by the upper bound
\begin{equation}
G_{i,j}^{(2)}(t)\leq\left(G_i^{(2)}(t)G_j^{(2)}(t)\right)^{\frac{1}{2}}
\end{equation}
We have introduced the single-time two-mode second-order
correlation functions
\begin{equation}
G_{i,j}^{(2)}(t)\equiv\langle\Phi(t)|\hat{a}_i^\dag\hat{a}_i\hat{a}^\dag_j\hat{a}_j|\Phi(t)\rangle
\end{equation}
and the single-time, single-mode, second-order correlation functions
\begin{equation}
G_j^{(2)}(t)\equiv\langle\Phi(t)|\hat{a}_j^\dag\hat{a}_j^\dag\hat{a}_j\hat{a}_j|\Phi(t)\rangle
\end{equation}

\begin{figure}
\includegraphics[width=3.4in,height=4.0in]{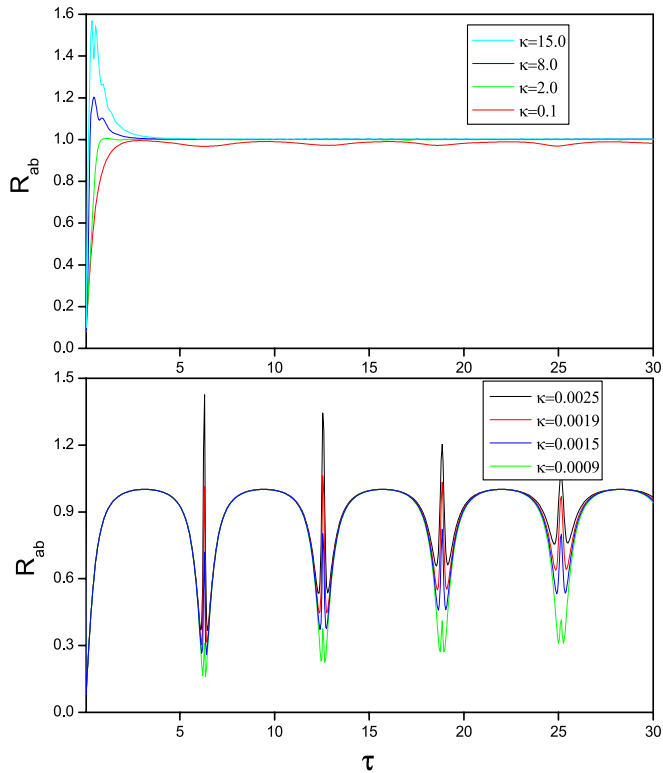}
 \caption{\label{cor} Normalized second-order cross-correlation function
 of two species in different wells. The parameter setting is same as FIG. (\ref{save}).}
\end{figure}

To examine the correlation, we calculate time dependence of the
second-order cross-correlation function between two species in
different wells
\begin{equation}
R_{a_L,b_R}^{(2)}\equiv\frac{G_{a_L,b_R}^{(2)}(t)}{\sqrt{G_{a_L}^{(2)}(t)G_{b_R}^{(2)}(t)}}
\end{equation}
The results are depicted in FIG. (\ref{cor}). These results
illustrate how the correlations between mode $a_L$ and $b_R$ do
violate  the classical Cauchy-Schwartz inequality when paired
tunnelling entanglement is prominent. The violation of second-order
two-mode correlation can be well understood by looking at the Von
Neumann entropy. Von Neumann entropy can quantify the degree of
entanglement of two species, but it can not be used to justify the
onset of entanglement. The entanglement of two species can only be
triggered when the correlation function is violated. Otherwise the
entropy comes from the superposition of all basis states. With the
help of second-order cross correlation function, one thus can
distinguish whether the system is entangled as well as measure the
degree of entanglement by Von Neumann entropy.

\section{discussion and conclusion}
Our systematic approach has shown, for the first time, the onset of
avoid level-crossing of two-component BECs system in a double well.
Different from the erratic level crossings of a single component
system \cite{wy}, we find the avoided level-crossing always exists
even for a intermediate system. As the number of particles of each
components increases, the avoided level-crossing happens during a
narrower range of $u=\kappa/\Omega$. What interesting is it shares
the similar character of the level repulsion  of an interacting
boson model \cite{arias}, where the system undergoes a transition
from SU(5) to SU(3). Other similar examples include the Floquet
spectra of one-dimensional driven systems \cite{latka}, in which the
level-crossings are induced by chaos. Note here our result is exact
the energy spectra. Mahmud et.al. \cite{mahmud} have studied the
double-well system using phase-space dynamics. Their results reveal
 even for a very small system (only 4-8 particles per well) the
classical chaos is manifest for a driven double-well BECs. However,
their Hamiltonian is slightly different from ours and their
calculation only demonstrates the merging of energy levels. The
system considered in this paper is much more complicated than the
one containing  single component BECs \cite{mahmud}. Although in the
highly degenerated region there are apparent signatures of
 the onset of entanglement \cite{ng,ng1}, in the avoided level-crossing region the
 dynamics has not been addressed clearly. It is not sure such
 crossing may relate to chaotic dynamics.

One may prefer Von Neumann entropy to quantity the entanglement
\cite{ng,ng1,chen1}. Our results imply entropy is not sufficient to
quantify whether the entanglement is triggered. However by
calculating the second-order crossing correlation function, the
dynamics can be obviously classified. In particular, the
second-order crossing correlation function will violate
Cauchy-Schwarz inequality when the system is entangled. Thus this
system is one of the candidates of many-particle entanglement of
BECs \cite{you,ng,ng1,chen1,micheli}. On the other hand, in the
strong interaction \cite{ng} of  the system (\ref{h2}), a slightly
difference between $\kappa_a=\kappa_b$ and $\kappa$ will destroy the
entanglement of two species. Thus it is hard to implement in a real
experiment.

In conclusion, we have studied the quantum dynamics of two-component
BECs in double-well thoroughly using the two-mode approximation
model \cite{milburn,steel,spekkens}. After performing a systemical
approach  \cite{wu1,wu,code} to obtain the eigenvalues of the
Hamiltonian (\ref{h2}), we find the energies become to degenerate in
the weak and strong interaction regime while undergo avoided
level-crossings, which may indicate some other dynamics. The quantum
dynamics studies demonstrate the self-trapping effect and
oscillation of the system. We show self-trapping of both the whole
system and individual component can happen with appropriate
conditions. We also study the quantum entanglement and correlation
in two extreme situations, i.e. weak interaction regime and strong
interaction regime. Our numerical experiments reveal that the onset
of entanglement of two species can be signed by the violation of
Cauchy-Schwarz inequality of second-order cross correlation
function. And the degree of entanglement is thus measured by Von
Neumann entropy.

Further perspectives of system (\ref{h2}) include what relates
avoided level-crossings  and whether it can cause certain kind of
chaos \cite{wy}. The quantum phase-space method
\cite{micheli,mahmud} may be a feasible candidate for the study of
these problems. And in a real system contains two-component BECs
\cite{stamper, hall,modugno}, $\kappa_a,\kappa_b,\kappa$ and
$\Omega_a,\Omega_b$ are actually different. What this influences
level, dynamics and entanglement may serve an interesting topic.

\begin{acknowledgments}
The work is supported in part by the NSF of China (Grant Nos.
60478029, 10125419, 10575040 and 90503010) and by the National
Fundamental Research Program of China, Grant No.2005CB724508.
\end{acknowledgments}

\end{document}